Entropy-Dominated Dissipation in Sapphire Shock-Compressed up to 400 GPa (4 Mbar)

W. J. Nellis,[1] G. I. Kanel,[2] S. V. Razorenov,[3] A.S. Savinykh,[3] and A. M. Rajendran[4*]

[1]Department of Physics, Harvard University, Cambridge, MA 02138, USA.

[2]Joint Institute for High Temperatures, Moscow, 125412 Russia.

[3]Institute of Problems of Chemical Physics, Chernogolovka, 142432 Russia

[4]U.S. Army Research Office, Rayleigh-Durham, NC 27709, USA.

Abstract

Sapphire (single-crystal $Al_2O_3$) is a representative Earth material and is used as a window and/or anvil in shock experiments. Pressure, for example, at the core-mantle boundary is about 130 gigapascals (GPa). Defects induced by 100-GPa shock waves cause sapphire to become opaque, which precludes measuring temperature with thermal radiance. We have measured wave profiles of sapphire crystals with several crystallographic orientations at shock pressures of 16, 23, and 86 GPa. At 23 GPa plastic-shock rise times are generally quite long (~100 ns) and their values depend sensitively on the direction of shock propagation in the crystal lattice. The long rise times are probably caused by the high strength of inter-atomic interactions in the ordered three-dimensional sapphire lattice. Our wave profiles and recent theoretical and laser-driven experimental results imply that sapphire disorders without significant shock heating up to about 400 GPa, above which $Al_2O_3$ is amorphous and must heat. This picture suggests that the characteristic shape of shock compression curves of many Earth materials at 100 GPa pressures is caused by a combination of entropy and temperature.

Sapphire is one of the most widely used, widely studied, and arguably one of the least understood materials under shock compression, particularly shock-induced optical effects from 15 to ~150 GPa. In 1970 Barker and Hollenbach found that 15 GPa is the maximum stress at which a *c*-cut sapphire crystal can be used as a window in a Velocity Interferometer for a Surface of Any Reflector (VISAR). The cause was "shock-induced luminescence and/or loss of transparency" [1]. Metallic fluid hydrogen has been synthesized at 140 GPa by a shock wave reverberating in liquid $H_2$ contained between two transparent *c*-cut sapphire crystals. The reverberating shock causes sapphire to become opaque, which precludes measurement of temperature from thermal spectra emitted from the dense fluid [2,3].

To try to understand and control the nature of shock-induced defects in sapphire and to see if there is a direction of shock propagation in the hexagonal crystal structure that remains more transparent to higher pressures than *c*-cut sapphire, we have measured temporal profiles of shock waves propagating in seven different directions in the sapphire lattice. The results were unexpected and led us to consider other experiments and recent theoretical calculations [4-7] to gain understanding of this unusual system. Remarkably, while shocked materials are generally considered to be at high temperatures, we discovered that strong sapphire is probably not shock-heated significantly until about 400 GPa. At lower pressures shocked sapphire disorders (entropy increases) continuously up to about 400 GPa at which sapphire is probably amorphous. Further dissipative losses at higher pressures must then heat amorphous $Al_2O_3$. The fact that sapphire is so strong causes the pressure of the transition from primarily entropic dissipation to the addition of significant shock heating to occur at such high pressure that the two effects are resolved and readily observed. This is the first time to our knowledge that effects of entropic and thermal dissipation have been resolved in shock flows [8].

To arrive at this conclusion it was necessary to consider our wave profile measurements in the context of previous work above 10 GPa. Hugoniots [9] of *c*-cut ($0001$), *a*-cut ($11\bar{2}0$), and *n*-cut ($11\bar{2}3$) sapphire crystals [10] and of polycrystalline alumina [11] were measured up to ~40 GPa with a rotating-mirror camera. The *a* axis is in the basal plane, the *c* axis is perpendicular to the basal plane, and the *n* axis makes an angle of $61^0$ with the *c* axis. Above the HELs [~15 GPa], the Hugoniots of these three crystal orientations and of alumina are essentially coincident, which was attributed to substantial loss of shear strength [10]. The calculated temperature rise of alumina at a shock stress of 30 GPa is small, only ~100 K [11]. Those Hugoniot data were analyzed assuming ideal elastic-plastic flow, though the rotating-mirror technique is weakly sensitive to time dependences of shock waves.

Urtiew found that shock-compressed *n*-cut sapphire starts to lose its transparency in the infrared (900 nm) at shock stresses in the range 100 to 130 GPa [12]. Kondo showed that between 16 and 85 GPa, optical spectra emitted from shocked sapphire are gray-body-like with an effective temperature of $5600 \pm 500$ K and an emissivity that increases up to 0.08 at 85 GPa [13]. This temperature is larger by an order of magnitude than estimated thermodynamically [11]. Kondo interpreted his spectra as thermal emission from heterogeneous hot spots whose areal density increases with shock stress. Hare et al took fast-framing photographs of *c*-, *a*-, and *r*-cut ($1\bar{1}02$) sapphire as shock waves of various amplitudes traversed those crystals. The *r* direction makes an angle of $58^0$ with the *c* axis. Hare et al observed the heterogeneous hot spots suggested by Kondo and their areal density depends on pressure and crystal orientation [14]. Local hot spots are indicative of local disorder (entropy).

The few shock-wave profiles that have been measured for sapphire have elastic-precursor decays and plastic waves with long rise times (~200 ns) at tens of GPa [15-18]. The cause of those long rise times was not explained. The response of quartz to shock compression is similar to that of sapphire [17].

Microstructures of sapphire recovered from high shock pressures were investigated for plastic flow and fracture. Recovered samples indicated that plastic deformation via slip and twinning is sensitive to the direction of shock propagation in the lattice and is largest for shocks traveling in the *r* direction relative to the *n* and *c* directions. At 24 GPa the volume fractions of twins are 0.18, 0.14, and 0.07 for shock propagation in the *r*, *n*, and *c* directions, respectively [19].

Hugoniot data from 20 to 140 GPa have been measured with plane-wave explosive systems [4] and with a two-stage gun from 80 to 340 GPa [5]. The fit to those Hugoniot data from 20 to 340 GPa is linear: $u_s=C+Su_p$, where $u_s$ is shock velocity, $u_p$ is particle velocity, C=0.874 cm/µsec, and slope S=0.957. Virtually all single-phase materials have S>1. Slope S<1.0 implies a phase transition in virtually all materials in which it has been observed. Thus, those Hugoniot data suggest a continuous phase transition in sapphire from 20 to 340 GPa, which begs the question as to what would cause S<1 over such a huge pressure range. Sapphire Hugoniot data also show scatter in the HEL near 30 GPa and a phase transition at 79 GPa [20]. A laser has been used to measure sapphire Hugoniot data from 1.1 to 1.9 TPa (19 Mbar) [6]. Those laser-driven data demonstrate the onset of significant thermal pressures and temperatures above 400 GPa.

To summarize, very few time-resolved VISAR profiles of sapphire crystals have been measured and those time histories are yet to be explained. Hugoniot data to 340 GPa is anomalous because it implies a continuous solid phase transition from 20 to 340 GPa! Calculated shock temperatures are low, ~1000 K at 100 GPa. Optical transparency had been shown to be heterogeneous, which is indicative of disorder/entropy, up to a shock pressure of 85 GPa and anisotropic, which is caused by the non-cubic crystal structure. Transparency up to 45 GPa has been shown to be correlated with the plastic deformation mechanisms of slip and twinning. Only a few directions of shock propagation in the hcp lattice have ever been investigated: along the *c*, *a*, *r* and *n* directions and *n* an *r* are almost coincident.

In order to systematically understand the ensemble of experimental data for solid hexagonal sapphire and to determine the direction of shock propagation in the lattice that is most transparent to the highest shock pressures, we have measured wave profiles for two thicknesses of seven sapphire crystal orientations at pressures of 16, 23, and 86 GPa. These pressures are (i) near the HEL, (ii) in the two-wave regime above the HEL, and (iii) just below the pressure at which the plastic wave overdrives the elastic wave. Two thicknesses were used to determine steadiness of each wave shape.

In this paper we report results only for *c*-cut and *r*-cut single crystals, which are sufficient to show that these wave profiles are quite unusual and to draw conclusions about the relative importance of entropic and thermal dissipation. These results also demonstrate the likely reason why the Hugoniots of strong materials, many of which are in the deep Earth, are virtually coincident with their cold compression curves at relatively low shock pressures and why these same materials have rapid upturns in shock pressure with compression at higher shock pressures [21].

Shock waves were generated by impact of a 2.0 mm-thick Al impactor plate at velocity $u_I$ onto the front surface of a 2.0 mm-thick Al base plate. The impactor plates were designed to preclude spall in them [22] and were accelerated with high explosives at the Institute of Problems of Chemical Physics. Impact velocities $u_I$ were $1.2 \pm 0.03$, $1.8 \pm 0.05$, and $5.2 \pm 0.1$ km/s, which produce peak stresses in sapphire of 16, 23, and 86 GPa. The Al base plate was backed by a sapphire disk, which in turn was backed by a LiF disk. An Al foil 7 microns thick separated the sapphire from the LiF. The laser beam of a VISAR was incident on axis on the rear surface of the target assembly. The 1-mm diameter beam passed through the transparent LiF disk and reflected off the Al foil at the sapphire/LiF interface. In this way velocity histories of sapphire/LiF interfaces were measured. Sapphire disks, obtained from Princeton Scientific Corporation, were 2.4 and 5.1 mm thick and 22 and 50 mm in diameter, respectively, thinner than used previously [1,10].

Use of two sample thicknesses for several orientation showed that wave shapes were nearly steady with run distance, as shown in Fig. 1 for *r*-cuts shocked to 23 GPa. Profiles of *r*-cuts are smoother with time than for *c*-cuts because the *r* direction has substantially more plastic deformation than along *c*. The elastic wave rises quickly (~ns) to a relatively steady amplitude, as does a conventional HEL. The rise time of the plastic wave is 100 ns. For *r*-cuts, HEL's (peak elastic stresses) were 13.3, 12.8, and 18.7 GPa for incident peak stresses of 16, 23, and 86 GPa, respectively.

Figure 2 shows wave profiles of 5.0 mm-thick *c*-cuts at three incident peak stresses. Fracture is significant in *c*-cuts, which causes profiles for c-cuts to be much more "noisy" in time than for *r*-cuts. For this reason profiles in Fig. 2 are each averages of two or three experiments. These averages are much smoother than individual profiles. Fracture will be discussed in a future publication. At 16 GPa the elastic shock is sharp and steady for 300 ns, after which it slowly decays. The corresponding elastic wave of *r*-cut sapphire at 16 GPa (not shown) starts to relax within 40 ns of reaching its maximum, suggesting elastic strength in the *r* direction is lower than in the *c* direction. At 23 GPa the elastic wave of *c*-cut rises quickly (~ns) and within a few ns starts to relax. The rise time of the second wave is ~300 ns. At an incident stress of 86 GPa, the elastic precursor first relaxes, which is followed by a plastic shock with a rise time of ~10 ns. Figure 2 suggests that above ~90 GPa only a single sharp (rise time<ns) plastic shock propagates. For *c*-cuts, HEL's were 16, 18, and 24 GPa for initial impact stresses of 16, 23, and 86 GPa, respectively.

The long rise times, ~$10^5$ ps, of plastic shocks in strong sapphire are 5 orders of magnitude longer than shock rise times of compressible fluids, ~1 ps in fluid Ar [23], for example. The long rise times are telling us that ~100 ns is required to break and/or reorient strong (~ev) bonds to reach maximum compression of sapphire. In shock-compressed liquid Ar the well depth for pair interactions is ~0.01 ev, 100 times smaller than ~ev bond strengths of $Al_2O_3$. At 50 GPa liquid Ar is shock-compressed to 2.2 fold in density and 14,000 K, and dissipative thermal energy is ~90% of shock energy [24]. At 50 GPa sapphire is shock-compressed to 1.1 fold in density and ~500 K, and dissipative thermal energy is only a few % of shock energy. These differences are huge and imply that the long rise times of shock waves in strong sapphire at tens of GPa are caused by the high strength of inter-atomic interactions in sapphire ordered in a three-dimensional lattice relative to a Vander Waals fluid with weak pair interactions. Above 50 GPa and 15,000 K, shock energy in fluid Ar is absorbed by thermal activation of electrons into a conduction band.

Calculated shock temperatures of sapphire are only ~1000 K at 100 GPa and the calculated 0-K isotherm up to 400 GPa is virtually coincident with Hugoniot measurements up to 340 GPa [7]. Because of the relatively low temperatures and thermal pressures, strong (~ev) inter-atomic interaction energies, and short experimental lifetimes, sapphire shocked up to 90 GPa is probably not in thermal equilibrium.

A non-equilibrium process that is expected to produce all these phenomena is bond breaking and reorientation that occurs continuously and statistically on a ~100 ns time scale. Shock temperatures are kept low from lowest shock pressures by absorption of ~ev of mechanical energy to break and/or reorient each bond. Energy absorbed mechanically is not available to heat the shocked sapphire, nor is this energy available to achieve thermal equilibrium. Rather, this mechanical process damages the lattices, which produces entropy.

The solid curves in Figure 3 are for the four phases that comprise the calculated 0-K isotherm of sapphire up to ~400 GPa calculated by Umemoto and Wentzcovitch [7]. The dots are Hugoniot data measured by Erskine [5] and by McQueen [4]. Theory and experiment are virtually coincident up to ~400 GPa, which even with taking into account error bars of ~1% or so in both, implies shock temperatures and thermal pressures are small compared to pressures on the 0-K isotherm. There are three oxide phases in the range of the fit obtained from those experiments. Given the ~100 ns experimental lifetimes, it is unlikely that shocked sapphire can anneal into these well-ordered crystalline phases. The good agreement between theory at 0 K and Hugoniot data means that sapphire is deforming plastically while changing its short-range order to the predicted structures. From Fig. 3 this process probably commences just below 90 GPa, the pressure of the first phase transition and the upper limit of our wave-profile measurements. A continuous phase transition is a source of disorder, that is, additional entropy. A continuous phase transition such as this over such a large pressure range is also a likely explanation of why slope S<1.0 over this entire pressure range and why shocked sapphire goes opaque above ~100 GPa. This is the first time to our knowledge that a calculated 0-K isotherm could be compared directly to measured Hugoniot data without thermal corrections over such large range of pressure.

At sufficiently high pressures, shock compression is expected to completely disorder sapphire to an amorphous state, which would maximize lattice entropy. At still higher shock pressures, dissipation would have to go primarily into temperature and thermal pressure. The pressure of this crossover can be estimated from laser-driven Hugoniot data up to 1.9 TPa in Fig. 4. Above 400 GPa, shock pressure increases dramatically with density, which is probably caused by the onset of thermal dissipation and increasing thermal pressure. From Fig. 3, 400 GPa is just above the calculated pressure of the $CaIrO_3$-type to $U_2S_3$-type transition, the highest-pressure transition predicted on the 0-K isotherm. The authors of [6] suggest that because of temperatures of several tens of 10,000 K, sapphire melts on the Hugoniot at ~500 GPa, above which liquid $Al_2O_3$ is a liquid semiconductor with a mobility gap of ~2 eV.

The high strength of sapphire is what enables recognition that below ~400 GPa, dissipation is primarily entropic and that above this pressure dissipation is the sum of maximum lattice entropy plus a thermal contribution that increases with increasing shock pressure. For materials with less strength than sapphire this clear separation of contributions has not been observed. At ~100 GPa significant entropic and thermal contributions to dissipation are both expected, which means the relatively sharp transition to significant thermal dissipation in Fig. 4 is not observed. This is the probable reason why many oxides have increases in slope of shock pressure with compression at ~100 GPa pressures [21]. In the case of $Gd_3Ga_5O_{12}$ (GGG), for example, the 300-K isotherm measured in a diamond anvil cell (DAC) [25] is nearly coincident with the measured Hugoniot [26] up to ~70 GPa, above which GGG becomes amorphous in a DAC. At a shock pressure of 120 GPa, GGG undergoes a transition to a material at ~5,000 K, which is probably a semiconducting liquid with a mobility gap of 3.1 ev [26], similar to $Al_2O_3$ [6].

# References

*Present address: Department of Mechanical Engineering. University of Mississippi, University, MS 38677

Figure Captions

Fig. 1  Wave profiles at sapphire/LiF interfaces after traveling through 2.4 and two 5.1-mm-thick *r*-cut sapphire disks and releasing into LiF. Incident shock stress into sapphire is 23 GPa for both, achieved by impact of Al on Al at 1.8 km/s.

Fig. 2  Wave profiles at sapphire/LiF interfaces after traveling through 5.0-mm-thick *c*-cut sapphire disks and releasing into LiF. Incident shock stresses into sapphire are 16, 23 and 86 GPa, achieved by impacts of Al onto Al at 1.2, 1.8, and 5.2 km/s. Profiles for 1.2 and 1.8 km/s are averages of 2 and 3 experiments, respectively. Profile for 5.2 km/s is average of two experiments with two different velocity-per-fringe VISAR constants.

Fig. 3  Solid curves are calculated pressure versus compression of 0-K isotherm of sapphire [**7**]. Four sapphire phases are calculated along 0-K isotherm. Horizontal dashed gray lines indicate pressures of transitions between phases. Dots are measured Hugoniot points [**4,5**].

Fig. 4  Measured Hugoniot data of sapphire from 1.1 to 1.9 TPa (19 Mbar) [**6**]. Data to 340 GPa [4,5] are in mutual agreement and coincident with 0-K isotherm (Fig. 3). Dashed curve is extrapolation of fit to Hugoniot data below 340 GPa. Data points at 1.1 to 1.9 TPa show onset of significant thermal pressure above 400 GPa.

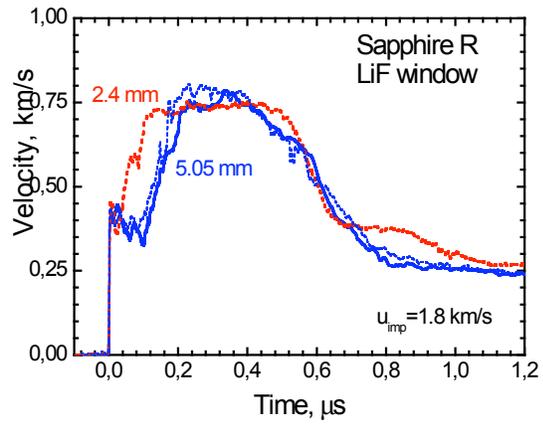

Fig. 1

Nellis et al

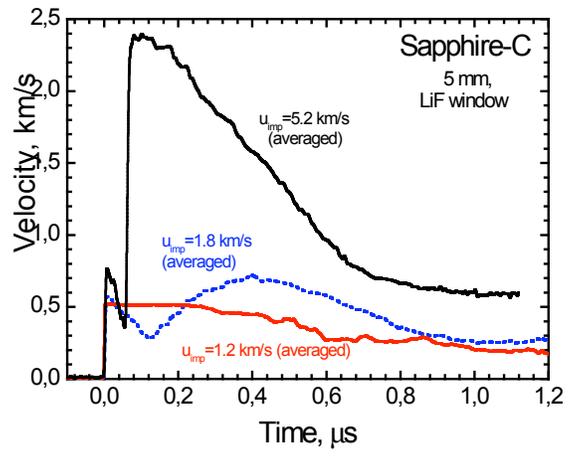

Fig. 2

Nellis et al

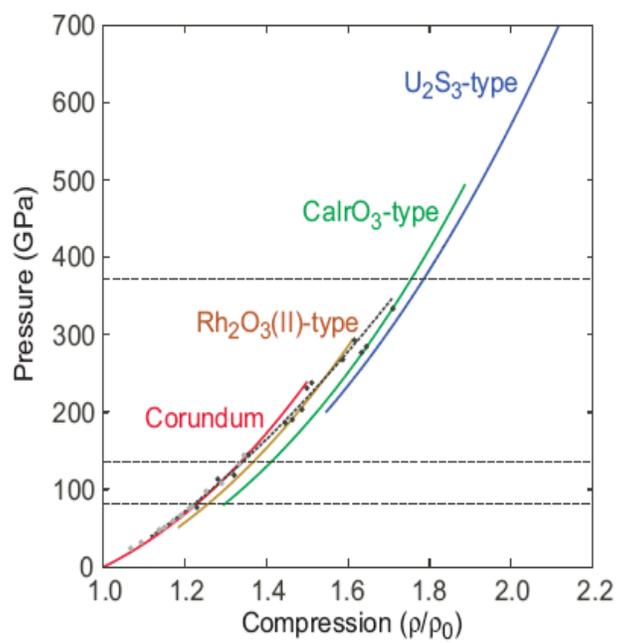

Fig. 3

Nellis et al

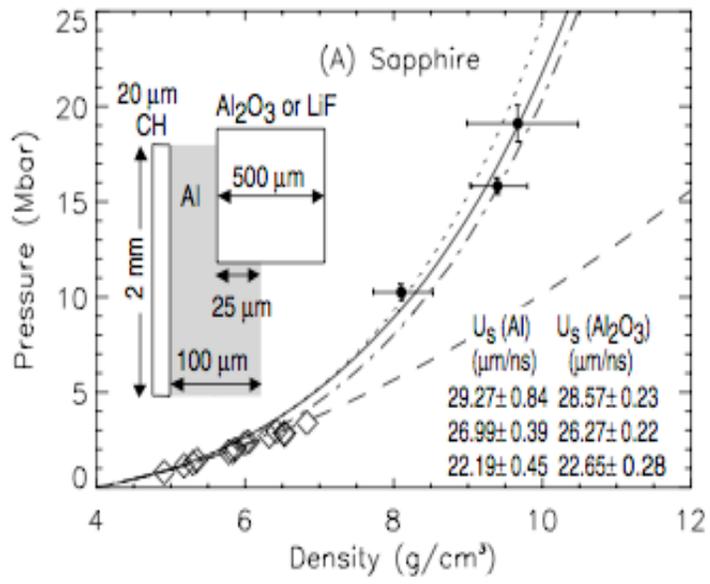

Figure 4

Nellis et al